\documentclass[aps,prd,twocolumn,showpacs,amsmath,nofootinbib]{revtex4-1}
\newcommand{\beqa}{\begin{eqnarray}}
\newcommand{\eeqa}{\end{eqnarray}}

\usepackage{graphicx}
\usepackage{epsfig}

\usepackage{graphicx}% Include figure files
\usepackage{dcolumn}% Align table columns on decimal point
\usepackage{bm}%

\begin{document}

\title{The contribution of ultracompact dark matter minihalos to
the isotropic radio background}

\author{Yupeng Yang$^{1,2,5}$} \email{yyp@chenwang.nju.edu.cn}
\author{Guilin Yang$^{1,5}$} 
\author{Xiaoyuan Huang$^{3}$} 
\author{Xuelei Chen$^{3}$} 
\author{Tan Lu$^{1,4,5}$}
\author{Hongshi Zong$^{1,5,6}$} \email{zonghs@chenwang.nju.edu.cn}

\affiliation{ $^1$Department of Physics, Nanjing University, Nanjing, 210093, China\\
$^2$Department of Astronomy, Nanjing University, Nanjing, 210093, China \\
$^3$National Astronomical Observatories, Chinese Academy of Sciences, Beijing, 100012, China\\
$^4$Purple Mountain Observatory, Chinese Academy of Sciences, Nanjing, 210008, China\\
$^5$Joint Center for Particle, Nuclear Physics and Cosmology, Nanjing, 210093, China\\
$^6$ State Key Laboratory of Theoretical Physics, Institute of Theoretical Physics, CAS, Beijing, 100190, China}

\begin{abstract}
The ultracompact minihalos could be formed during the earlier epoch of 
the universe. The dark matter annihilation within them is very
strong due to the steep density profile, $\rho \sim r^{-2.25}$. The high energy electrons and positrons   
from the dark matter annihilation can inverse Compton 
scatter (ICS) with the background photons, such as CMB photons, to acquire higher 
energy. On the other hand, the synchrotron radiation can also 
be produced when they meet the magnetic field. 
In this paper, we study the signals from the UCMHs due to 
the dark matter annihilation for the radio, X-ray and $\gamma$-ray band. 
We found that for the radio emission the UCMHs can provide one kind of 
source for the radio excess observed by ARCADE 2. 
 But the X-ray signals due to the ICS effect or the $\gamma$-ray 
signals mainly due to the prompt emission from dark matter would exceed
the present observations, such as Fermi, COMPTEL and CHANDRA. We found that 
the strongest limits on the fraction of UCMHs come from the X-ray observations 
and the constraints from the radio data are the weakest.
\end{abstract}

%\keywords{
%ltracompact minihalos, dark matter, annihilation, radio, X-ray, $\gamma$-ray}
\maketitle

\section{Introduction}
It has been confirmed by many observations and 
theoretical research that the present
structures of our universe come from the earlier density perturbations 
$\delta \rho/\rho \sim 10^{-5}$.
It was proposed that at the earlier epoch, if the density perturbations 
are larger than $\sim 0.3$, the 
primordial black holes (PBHs) would be formed \cite{y1,y2}. These
large perturbations cannot be achieved within the current popular 
theory which has predicted a scale invariant spectrum 
$P(k) \propto k^{1-n}$, and the present observations give 
$n = 0.968 \pm 0.012$~\cite{wmap7}. 
\footnote{The new results from the WMAP-9 year data show that 
there is a tilt in the primordial spectrum~\cite{wmap9}.} 
So the large density perturbations can 
only be produced under some special conditions, such as 
the cosmological phase transitions or the feature at some scale of 
inflation potential~\cite{y6,y3,y4,y5,y7}. In Ref. \cite{y8}, the authors proposed that if 
the density perturbations during the radiation dominated epoch are less than 0.3 
but larger than $10^{-3}$, one new kind of dark matter structures named 
ultracompact minihalos (UCMHs) would be formed. 
Because the density perturbations needed by the formation of UCMHs are smaller than PBHs, 
there are larger probability to form these new objects. 
After the formation of UCMHs, they will accrete the matter onto them through 
the radial infall. 
Due to the steep density profile and the early formation time of UCMHs, 
it is excepted that compared with the 
standard dark matter halos,
these compact objects would have some different 
and notable effect on the cosmological evolution. In Refs.~\cite{y10,y9,y11}, 
the authors discussed the influence of the UCMHs on the CMB
due to the dark matter annihilation within them and obtained the constraints 
on the abundance of UCMHs. The $\gamma$-ray flux from 
the UCMHs are also studied by several authors~\cite{y12,y13,y14}. 
Besides the $\gamma$-ray, the electrons and positrons can also 
be produced from the UCMHs due to the dark matter annihilation. 
The synchrotron radiation will be produced due to the meeting between 
these high energy charged particles with the intergalactic magnetic fields. 
These radio signals would contribute to the cosmological background. 
For the cosmological radio background, the dominated contributions 
at frequencies above several GHz are from the cosmic microwave background (CMB). 
At the lower frequencies, the main contributions come from the 
extra-galactic radio sources which have been detected by the current observations~\cite{radio_1,radio_2}.
Recently, the radio flux excess with respect to the total contributions from the detected 
extra-galactic radio sources 
in the lower frequency region, $\nu \lesssim$ 10GHz, was
observed by ARCADE 2~\cite{y15}. The final results are obtained by analyzing the data of
the ARCADE-2 collaboration
and older surveys at lower
frequency observations~\cite{radio1,radio2,radio3,radio4}.
These observations cannot be explained 
even when the unresolved astrophysical objects are included~\cite{y16,y17}. 
The authors of Refs.~\cite{y18,y19} found that the dark matter annihilation 
within the dark matter halos  
can fit the observations. On the other hand, these high energy electrons and positrons
 can inverse Compton scatter with the CMB photons 
into the X-ray or $\gamma$-ray band~\cite{y18,y19,y20}. In this work, 
we studied the radio signals produced by the dark matter annihilation 
within the UCMHs and the corresponding X-ray and $\gamma$-ray signals. 
Using the observational data from Fermi, COMPTL, CHANDRA and the ARCADE, 
the limits on the fraction of UCMHs are obtained.  

This paper is organized as follows. The basic characteristic of the UCMHs are 
discussed in Sec. II. In Sec. III we investigate the radio, X-ray and 
$\gamma$-ray signals from the UCMHs due to the dark matter annihilation and 
obtain the constraints on the fraction of UCMHs using these band observations. 
The conclusions and discussions are presented in Sec. IV

\section{The formation and growth of UCMHs.}
If the density perturbations during the radiation dominate epoch 
satisfy the condition $10^{-3} \lesssim \delta \rho/\rho \lesssim 0.3$, 
one new kind of structures named ultracompact minihalos 
would be formed. In fact, the minimal value of the density perturbations 
depends on the time of horizon entry 
of the scale (for more detailed discussions one can see Ref.~\cite{y21}).
After the formation, the mass of UCMHs grows slowly because of 
the Meszaros effect until after the matter-radiation equality. The evolution of the
mass has the form~\cite{y12} 

\begin{equation}
\label{Mh}
M_\mathrm{UCMHs}(z) = \delta m \left(\frac{1 + z_\mathrm{eq}}{1+z}\right),
\end{equation}
where $\delta m$ is the mass within the scale of perturbations 
and it is different at different redshift. 
The density profile of UCMHs are obtained through 
the simulations~\cite{y8}, $\rho \propto r^{-9/4}$. 
%Of cause, there are 
%some different between the inter and outer part~\cite{}. Here we 
%use the above one. 
So, the specific form of the density profile can be written as 

\begin{equation}
\label{density}
\rho_\mathrm{UCMHs}(r,z) = \frac{3f_\chi M_\mathrm{UCMHs}(z)}{16\pi R_\mathrm{UCMHs}(z)^\frac{3}{4}r^\frac{9}{4}},
\end{equation}
where ${R_\mathrm{UCMHs}(z)} = 
0.019\left(\frac{1000}{z+1}\right)\left(\frac{M_\mathrm{UCMHs}(z)}
{\mathrm{M}_\odot}\right)^\frac{1}{3} \mathrm{pc}$ and $f_{\chi} = \frac{\Omega_{DM}}{\Omega_b+\Omega_{DM}} = 0.83$~\cite{wmap7} 
is the dark matter fraction. 
After $z \sim 100$, the structure formation will dominate, so 
in this work we adopt the assumption that the UCMHs stop 
growing at $z \sim 10$~\cite{y12,y11,y21}. On the other hand, since the dark matter 
annihilation will soften the central density of UCMHs, 
there is a maximal density $\rho_{max}$ at time $t$ for the formation 
time of UCMHs $t_i$, $\rho_{max}(r_{min}) =  m_{\chi}/\langle \sigma v \rangle (t-t_i)$. 
Following the previous works~\cite{y10,y9,y11,y14,y12,y13}, we truncate the density profile at $r=r_{min}$ and take 
the density within this radius as a constant, $\rho(r<r_{min}) = \rho_{max}$.

\section{The multi-band signals from dark matter annihilation within the UCMHs.}

As the essential component of the cosmos,
dark matter has been confirmed by 
many observations. But its nature remains unknown. Now there are many 
dark matter models, and the much studied one is the 
weakly interacting massive particles (WIMPs)~\cite{y22,y23}. 
One of the important models within the WIMP is the neutralino. 
According to the theory, these particles can annihilate into 
the standard particles, such as photon, electron and positron. 
The multi-band signals produced by these high energy particles has been 
studied as the clue of looking for the dark matter~\cite{y24,y25}. 
Recently, the ARCADE 2 (Absolute Radiometer for Cosmology, 
Astrophysics and Diffuse Emission) released the results of the radio observations 
and found the excess at the lower frequency~\cite{y15}. These results 
cannot be explained by the classical astrophysical sources. 
In Refs.~\cite{y18,y19}, the authors suggested that 
the classical dark matter halos due to the dark matter annihilation may be one kind of the 
sources for the excess. Besides the radio emission, the corresponding 
X-ray and $\gamma$-ray signals can also be produced through 
the dark matter annihilation~\cite{y18,y19,y20}. The authors of Ref.~\cite{y20} 
found that only the dark matter models in which the mass is smaller and 
the dominating annihilation channel is the lepton channel can satisfy 
all band observations. 

Compared with the standard dark matter halos, the density profile of UCMHs 
is steeper and the formation time is earlier. So it is expected
that these objects can have significant contributions to the 
cosmological background~\cite{y13}. 
In this work, we study whether the UCMHs can provide the sources for 
the radio excess. We also study whether the 
corresponding X-ray and $\gamma$-ray emission is consistent with 
other present observations, such as CHANDRA, COMPTEL and Fermi. 

\begin{figure}
%\begin{center}
\epsfig{file=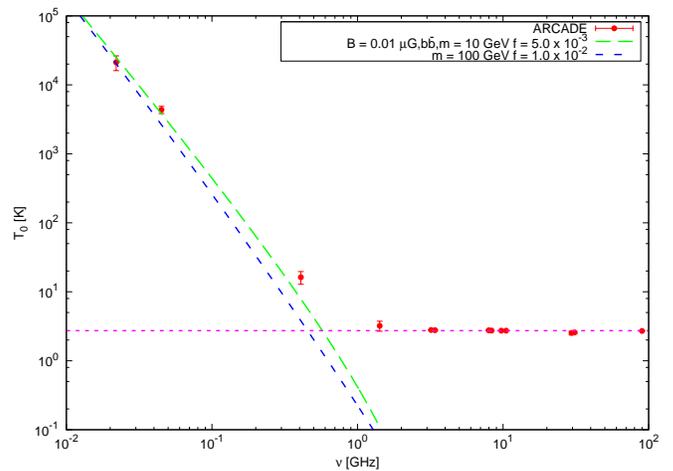,width=0.5\textwidth}
\caption{The radio signals from UCMHs for the $b \bar b$ channel is shown. 
The mass of dark matter is 10GeV and 100GeV, and the corresponding 
best fitting value of the UCMHs fraction is 
$f = 5.0 \times 10^{-3}$ and $1.0 \times 10^{-2}$, respectively. 
We have set the intergalactic magnetic field  $B = 0.01\mu G$ and 
the annihilation cross section of dark matter $\left\langle \sigma v
\right\rangle = 3.0 \times 10^{-26}cm^{-3}s^{-1}$. The horizontal line corresponds 
to the CMB temperature: $T_0 = 2.73 K$.} 
%\end{center}
\label{radio_bb}
\end{figure}

\begin{figure}
\epsfig{file=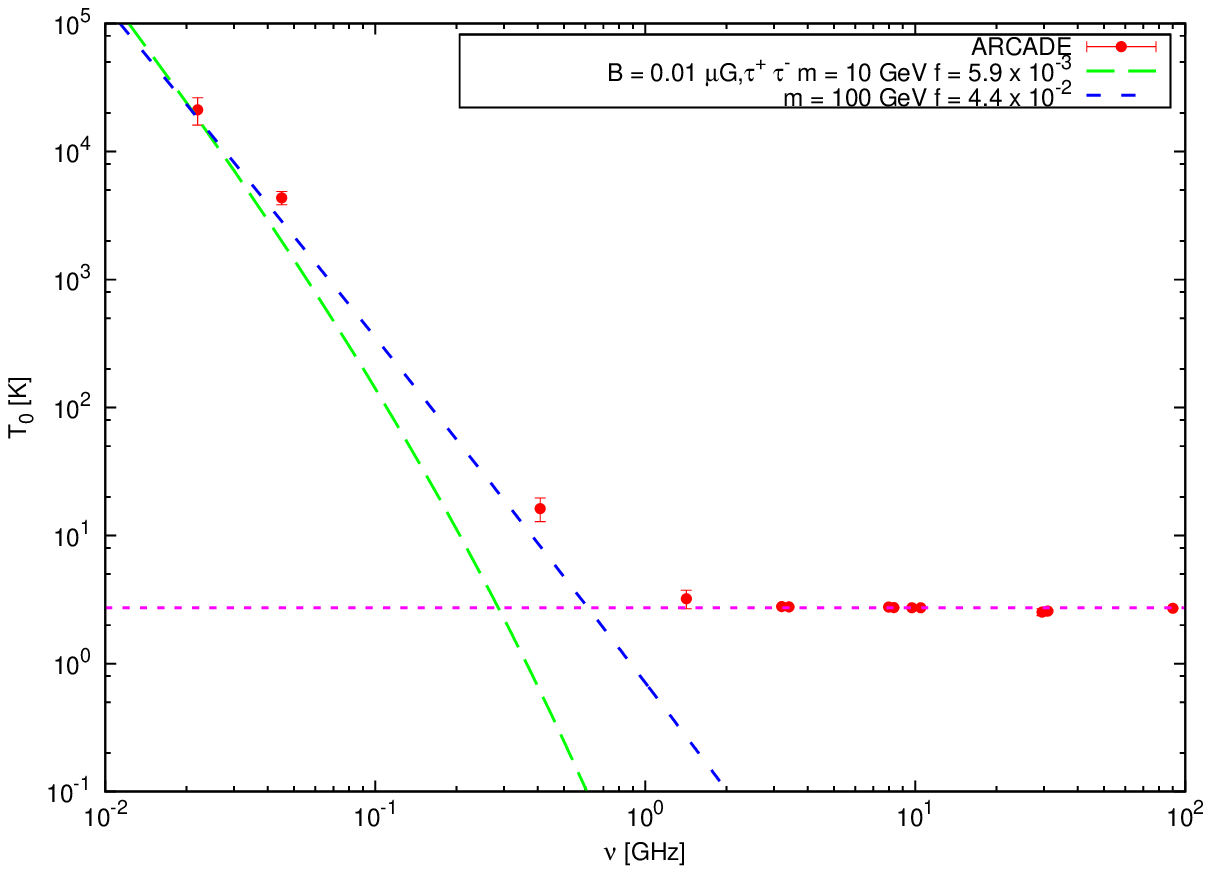,width=0.5\textwidth}
\caption{The radio signals from UCMHs for the 
$\tau^+ \tau^-$ channel. The other parameters are the same as Fig.~\ref{radio_bb}.}
\label{radio_tau}
\end{figure}

\begin{figure}%[t]{0.45\linewidth}
\epsfig{file=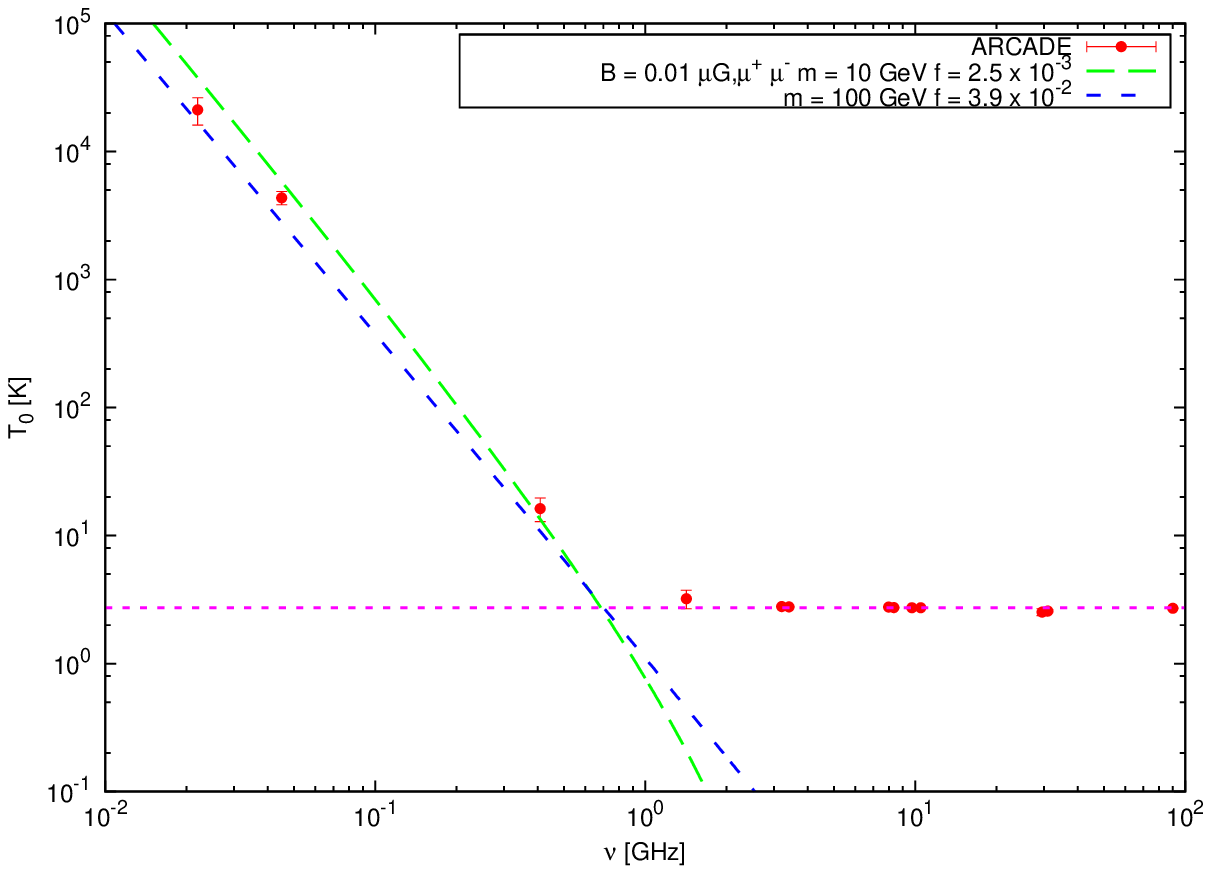,width=0.5\textwidth}
\caption{The radio signals from UCMHs for the 
$\mu^+\mu^-$ channel. The other parameters are the same as Fig.~\ref{radio_bb}.}
\label{radio_mu}
\end{figure}

\begin{figure}%[h]
\epsfig{file=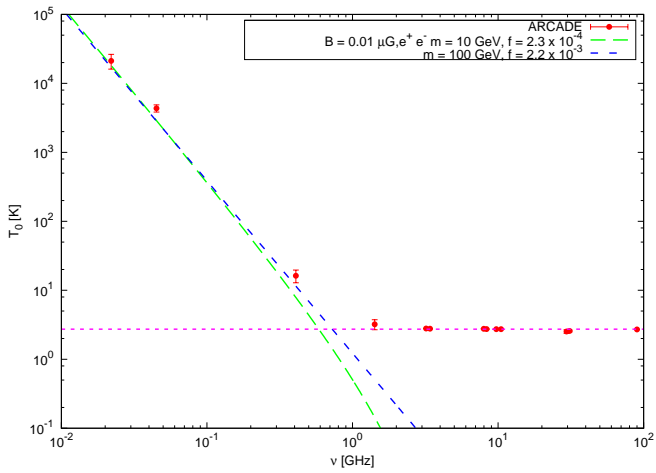,width=0.5\textwidth}
\caption{The radio signals from UCMHs for the 
$e^+e^-$ channel. The other parameters are the same as Fig.~\ref{radio_bb}.}
\label{radio_ee}
\end{figure}

The signals from the UCMHs can be written as~\cite{y18,y19,ics}:

\beqa
%\nu\,I_\nu
F_{\nu}=\frac{c\,\nu}{4\pi}\int dz\frac{e^{-\tau(z)}}{(1+z)\,H(z)}
\int dM \left(\frac{dn}{dM}\right)_\mathrm{UCMHs} \nonumber\\
\times \mathcal{L}(E,z,M)
\label{eq:intgenfirst}
\eeqa
%where we have choose the mass of the objects as the parameter. 
where $\tau$ is the optical depth and $\mathcal{L}$ is the luminosity of UCMH.
$\mathcal{L}$ depends on the redshift and the mass of UCMHs, 

\beqa
\mathcal{L}=\frac{\langle \sigma v \rangle}{2m_{\chi}^2}\times 
\int \rho^2_\mathrm{UCMHs}(r,z)d^{3}r \times \nonumber\\ 
\int P(r,E,E_\nu)&&\times
\left(\frac{1}{\dot E}\int\frac{dN_e}{dE'} dE'\right)dE
\label{eq:intgenlum}
\eeqa
where $dN_e/dE$ is the energy spectrum of the electron which can be obtained from 
the public code DarkSUSY~\footnote{http://www.physto.se/~edsjo/darksusy/},
$\dot E$ is the energy loss rate $\dot E = 3 \times 10^{-17}(1+z)^4 E^2 
\mathrm{GeV s^{-1}}$~\cite{ics} and
$P$ is the emission power. For the synchrotron case and the inverse Compton process $P$
can be written as

\begin{equation}
 P_{syn} (r,E,\nu)= \frac{\sqrt{3}\,e^3}{m_e c^2} \,B(r) F(\nu/\nu_c),
\end{equation}

\begin{equation}
P_{IC}(r,E,E_{\nu}) = c\,E_{\nu} \int d\epsilon\, n_\gamma(\epsilon,r)\,\sigma(\epsilon,E_{\nu},E)
\label{eq:PIC},
\end{equation}
where $\nu_c \equiv \frac{3}{4\pi} \frac{ce}{{({m_{e}c^2})}^3} B(r) E^2$, 
$m_e$ is the electron mass, and $B(r)$ is the strength of the magnetic field,  
which is usually a function of the position. In this work, 
we consider the cosmological contributions from the UCMHs, so the 
dependence of the magnetic field on the position is not important~\cite{ics}. 
After the formation of UCMHs and with the evolution of the universe, 
the UCMHs would be attracted into the classical dark matter halos due to the tidal force 
after the redshift $z \sim 100$. Therefore, the UCMHs would be distributed in 
some specific form, e.g. the NFW distribution. 
In this work, for simplicity, we assume that 
the UCMHs are distributed uniformly in the intergalactic space of the universe. 
The case that UCMHs are within the galaxy or cluster are also discussed in the next section.
Therefore, we use the intergalactic magnetic field value and they have been determined by several observations~\cite{magnetic_1,magnetic_2}. 
However, there are much uncertainty about these results. 
In this work, we take the conservative value of the intergalactic magnetic field 
as $B = 0.01 \mu G$ and our results can be applied to the case of other values of the magnetic field.
For the mass function of UCMHs, $dn/dM$, 
following the previous works~\cite{y10,y9,y11,y14,y12,y13}, 
we use the delta form $dn/dM \sim \delta(M-M_\mathrm{UCMHs})$. 
We also assume that UCMHs do not merge between them during their evolution. 
Therefore, we can define the fraction of UCMHs at present~\cite{y13} 
$f_\mathrm{UCMHs} = \rho_\mathrm{UCMHs}/\rho_{c}$, where $\rho_{c}$ is the critical density.
Using these assumptions and definition, Eq.~\ref{eq:intgen} can be rewrite as

\beqa
F_{\nu}=\frac{c\,\nu}{4\pi}\frac{f_\mathrm{UCMHs}\rho_{c}}{M_\mathrm{UCMHs,0}}\int dz\frac{(1+z)^2e^{-\tau(z)}}{H(z)}
\mathcal{L}(E,z,M)
\label{eq:intgen}
\eeqa
From Eq.~\ref{eq:intgenlum} it can be seen that the luminosity of UCMHs 
is proportional to $\delta m$, $\mathcal{L} \propto 
\int \rho^2_\mathrm{UCMHs}(r,z)d^{3}r \propto \delta m$,
\footnote{We have taken the lower limit of integration of the density profile of UCMHs 
as $r_{min}$ and $\rho_{\mathrm{UCMHs}}(r_{min}) = m_{\chi}/\langle \sigma v \rangle t$, 
so $r_{min} \propto \delta m^{1/3}$. Therefore, the luminosity $\mathcal{L}$ is proportional to $\delta m^{-1/2} \times 
\delta m^{3/2} = \delta m$. The latter factor $\delta m^{3/2}$ is from the other part of Eq.~\ref{density} 
except for $r^{-9/4}$.} 
so the final signals $F_\nu$ in Eq.~\ref{eq:intgen} is independent of 
$\delta m$, $F_\nu \propto \frac{1}{\delta m} \times \delta m$. 
In this work, we consider four annihilation channels: 
$b \bar b, \tau^+\tau^-, \mu^+\mu^-,$ and $e^+e^-$. 
The $b \bar b$ and $\tau^+\tau^-$ channel are the typical annihilation channels 
of the neutralino. The $\mu^+\mu^-$ and $e^+e^-$ channels are favored by 
the recent observations of positrons fraction: 
PAMELA~\cite{pamela} and ATIC~\cite{atic}. We fix the value of the dark matter mass and 
the annihilation cross section and adjust the fraction of the UCMHs to obtain the 
best fitting value for the radio data. 
In Figs.~\ref{radio_bb} and ~\ref{radio_tau}, we show the radio signals from the UMCHs 
due to the neutralino annihilation for the 
best fitting value of the fraction for the two typical channels. 
For the $b \bar b$ channel, the best values of the UCMHs fraction 
are $f = 5.0 \times 10^{-3}, 1.0 \times 10^{-2}$ for the dark matter 
mass 10GeV and 100GeV, respectively. For the $\tau^+\tau^-$ channel, the 
fraction of UCMHs are $f = 4.4 \times 10^{-2}$ and $5.9 \times 10^{-3}$, 
respectively. 
In Figs.~\ref{radio_mu} and ~\ref{radio_ee}, the results for lepton channels 
$\mu^+\mu^-$ and $e^+e^-$ are shown. The fraction of UCMHs are 
$f = 2.5 \times 10^{-3}, 3.9 \times 10^{-2}$ and 
$f = 2.3 \times 10^{-4}, 2.2 \times 10^{-3}$ for $m_\chi$ = 10 GeV and 100 GeV, respectively. 
For all these plots, we have set the dark matter annihilation cross section 
$\left\langle \sigma v \right\rangle = 3 \times 10^{-26} \mathrm{cm^{-3} s^{-1}}$.


\begin{figure}
%\begin{center}
\epsfig{file=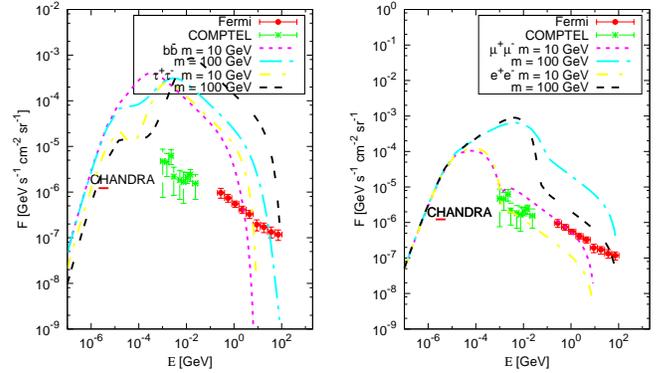,width=0.5\textwidth}
  \caption{The ICS and prompt emission from the UCMHs due to the dark 
matter annihilation. 
Left panel: $b \bar b$ and $\tau^+ \tau^-$ channels, 
the prompt emission are also included. Right panel: $\mu^+\mu^-$ and $e^+e^-$ channels. Here the final state radiation are also included.  
The data of Fermi, COMPTEL and CHANDRA are also shown.}
%\end{center}
\label{ics_ch}
\end{figure}

\begin{figure}
%\begin{center}
\epsfig{file=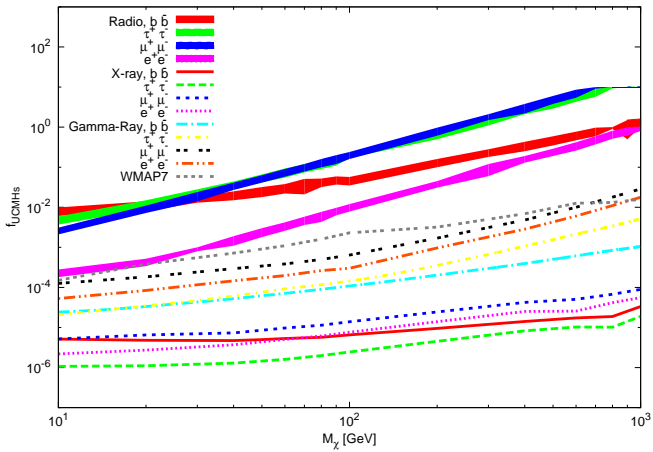,width=0.5\textwidth}
  \caption{Constraints on the abundance of UCMHs from the ARCADE, Fermi, COMPTEL 
data and CHANDRA observations 
for different channels, $b\bar b$,$\tau^+ \tau^-$, $\mu^+\mu^-$ and $e^+ e^-$, 
used in this work. For all band observations, the $95\%$ confidence regions 
and limits are shown.
Here we have fixed the cross section $\left\langle \sigma v \right\rangle 
= 3.0 \times 10^{-26} cm^{-3}s^{-1}$ for the dark matter annihilation.
The constraints from the WMAP7 data are also shown~\cite{y13}.}
%\end{center}
\label{radio_fit}
\end{figure}

The electrons and positrons which contribute to the radio signals 
due to the dark matter annihilation will also inverse Compton scatter
with the CMB 
photons to make them go into higher energy band, such as X-ray or soft $\gamma$-ray. 
On the other hand, the prompt emission can also 
contribute to the extragalactic $\gamma$-ray background~\cite{ics}. 
Although the dark matter model mentioned above can fit the radio 
data, these models must be consistent with other observations, 
such as X-ray and $\gamma$-ray~\cite{fermi,comptel,y18,y19,y20}. 
In Fig.~\ref{ics_ch} we show the signals of X-ray 
and $\gamma$-ray band from the UCMHs for the dark matter models  
which have been used to explain the radio excess. 
From this figure, we can see that the X-ray and $\gamma$-ray signals from those dark matter models 
are not consistent with the present observations: Fermi~\cite{fermi}, 
COMPTEL~\cite{comptel} and CHANDRA~\cite{chandra}. 
One of the important reasons is that the formation time of UCMHs is earlier, so
the signals from the higher redshift can also have significant contribution 
to, for example, the soft $\gamma$-ray background. In order to be consistent with the
observations, the constraints on the fraction of UCMHs can be obtained using these data. 
Firstly, in Fig.~\ref{radio_fit} we show the constraints on the 
UCMHs fraction from the radio excess data. 
On the other hand, the constraints on the UCMHs can be obtained from the CMB data.   
The product of the dark matter annihilation, such as photons, 
electrons and positrons, will interact with other particles 
existing in the universe. These effect will have impact on 
the ionization. The evolution of the electron fraction including the dark matter 
annihilation can be written as~\cite{xuelei}

\begin{equation}
\label{reion}
\frac{dx_e}{dz} = \frac{1}{(1+z)H(z)}[R_s(z)-I_s(z)-I_{DM}(z)],
\end{equation}
where $R_s$ and $I_s$ are the standard recombination rate and ionization rate, respectively, and
$I_{DM}$ is the ionization rate from the dark matter annihilation. In this work, 
we consider the contributions from the UCMHs. 
The change of the evolution of the ionization can impact the power spectrum of CMB. 
So the parameters such as the dark matter mass and the annihilation 
cross section can be constrained by the CMB 
observations. In Refs.~\cite{y9,y13}, the authors have investigated the 
impact of UCMHs on the CMB and obtained the constraints on the fraction of UCMHs 
using the WMAP7 data. In Fig.~\ref{radio_fit} we also show 
the constraints from WMAP7 data \cite{y13}. In order to be consistent 
with the Fermi, COMPTEL and CHANDRA observations, the limits for the fraction 
of UCMHs from these data are also given in the figure. 
From this figure one can see that the strongest limits come from the X-ray 
observations and the constraints from the radio data are the weakest. 
\footnote{The results from the WMAP7 data are independent of
the dark matter annihilation channels.}

\section{Discussion and conclusion}

Due to the steep density profile and the early formation time of UCMHs, 
the dark matter annihilation within them will have potential significance 
as the astrophysical sources. The recent results of ARCADE 2 
show that there are excess of radio signals for the frequency $\nu \lesssim 10$GHz. 
These results cannot be explained by the classical astrophysical objects 
even when the unresolved sources are also included. One of the possible explanations is 
the dark matter annihilation within the halos. In this work,  
we have studied the contributions of UCMHs to these observations. 
The dark matter particle mass considered here 
spread from 10GeV to 1TeV, and this range is favored 
by the direct or indirect observations. For the dark matter annihilation channels 
considered by us, $b \bar b$, $\tau^+ \tau^-$, $\mu^+ \mu^-$ and $e^+ e^-$, 
the radio signals from the UCMHs can fit the ARCADE data for the different 
parameters. In this work, we use the radio data to obtain the constraints on the 
fraction of UCMHs. On the other hand, the high energy electrons and positrons can inverse Compton
scatter with the background photons, such as CMB photons, to make them go into the X-ray or 
$\gamma$-ray band. We studies these signals from the UCMHs using the models 
which can fit the ARCADE data. We found that the X-ray or $\gamma$-ray signals 
exceed the present observations, such as CHANDRA, COMPTEL and Fermi data. 
Using these data, we also obtained the limits on the fraction of UCMHs. 
Compared with the results from the radio data, the constraints are stronger, 
especially for the X-ray observations, and 
the strongest limit is $f_\mathrm{UCMHs} \sim 10^{-6}$ for the dark matter 
mass $m_\chi = 10 \mathrm{GeV}$. 
In this work, we do not consider the standard dark matter halos. In
Refs.~\cite{y18,y19,y20}, their contributions have been discussed in detail. The authors 
of~\cite{y20} found that in order to fit the radio data and be consistent with 
the $\gamma$-ray observations, only those dark matter 
models which annihilate into the $\mu^+ \mu^-$ or $e^+ e^-$ and the mass is in the 
range of $5\sim50$GeV can satisfy all band observations.
Moreover, considering the uncertainties of the magnetic field 
and the density profile of halos, the dark matter annihilation cross section 
is in the range, $\langle \sigma v \rangle \sim (0.4 - 30) \times 10^{-26} 
\mathrm{cm^{-3}s^{-1}}$. So, if the standard dark matter halos are included, the constraints 
on the fraction of UCMHs obtained in this work would be changed and the details 
depend on the parameters of dark matter, the density profile of halos and the magnetic field. 
On the other hand, we have considered the homogeneous distribution of UCMHs and 
used the intergalactic field. The value of the magnetic field is smaller 
than the case of galaxy. If the UCMHs are present in the galaxy 
and the cluster, the constraints on the fraction of UCMHs will become stronger. 

In summary, we have studied the radio, X-ray and $\gamma$-ray 
signals from the UCMHs due to the dark matter annihilation. 
 We found that under the reasonable assumptions of the related parameters, 
the UCMHs which fit the radio excess observations 
are not consistent with the other band data. The limits on the fraction of UCMHs 
are obtained from all band observations, ARCADE 2, COMPTEL, CHANDRA and Fermi data 
and the strongest constraints come from the X-ray data.  

\section*{Acknowledgments}

Yang Yu-Peng thank Sun Weimin for improving the
manuscript and thank Huang Feng, Yuan Qiang and Feng Lei for very useful suggestions and discussions. 
This work is supported in part by the National Natural Science Foundation of China 
(under GrantNos 10935001, 11075075 and 11275097) and the Research Fund for the Doctoral Program of Higher
Education (under Grant No 2012009111002).

\providecommand{\href}[2]{#2}\begingroup\raggedright\begin
{thebibliography}{10}

\bibitem{y1}
Y.~B.~{Zeldovich}, I.~D.~{Novikov},      
{\em Sov. Astron.} {\bf 10} (1961) 602.

\bibitem{y2}
A.~E.~{Guzman}, J.~{May}, H.~{Alvarez}, K.~{Maeda},
%  {\it Gravitationally collapsed objects of very low mass}, 
{\em MNRAS.} {\bf 152} (1971) 75.

\bibitem{wmap7}
E.~{Komatsu} et al,    
%  {\it Seven-year Wilkinson Microwave Anisotropy Probe (WMAP) Observations: Cosmological Interpretation}, 
{\em Astrophys. J. Suppl.} {\bf 192} (2011) 192.
%[\href{http://xxx.lanl.gov/abs/1001.4538}{{\tt 1001.4538}}].
[arXiv:1001.4538]

\bibitem{wmap9} G. Hinshaw et al, arXiv:1212.5226 

\bibitem{y7}
B.~J.~{Carr}, K.~{Kohri}, Y.~{Sendouda}, J.~{Yokoyama}, 
%  {\it New cosmological constraints on primordial black holes}, 
{\em Phys. Rev.} {\bf D81} (2010) 104019.
%[\href{http://xxx.lanl.gov/abs/astro-ph/0912.5297}{{\tt 0912.5297}}].
[arXiv:0912.5297]

\bibitem{y3}
P.~{Ivanov}, P.~{Naselsky}, I.~{Novikov},   
%  {\it Inflation and primordial black holes as dark matter}, 
{\em Phys. Rev.} {\bf D50} (1994) 7173.

\bibitem{y5}
R.~{Saito}, J.~{Yokoyama}, R.~{Nagata},  
%  {\it Single-field inflation, anomalous enhancement of superhorizon fluctuations, 
%and non-Gaussianity in primordial black hole formation }, 
{\em JCAP} {\bf 06} (2008) 024.
%[\href{http://xxx.lanl.gov/abs/astro-ph/0804.3470}{{\tt 0804.3470}}].
[arXiv:0804.3470]

\bibitem{y6}
C.~{Schmid}, D.~J.~{Schwarz}, P.~{Widerin},   
%  {\it Peaks above the Harrison-Zel'dovich Spectrum due to the Quark-Gluon to Hadron Transition}, 
{\em Phys. Rev. Lett.} {\bf 78} (1997) 791.
%[\href{http://xxx.lanl.gov/abs/astro-ph/9606125}{{\tt astro-ph/9606125}}].
[astro-ph/9606125]

\bibitem{y4}
J.~{Yokoyama},      
%  {\it Formation of primordial black holes in the inflationary universe}, 
{\em Phys. Rep.} {\bf 307} (1998) 133.

\bibitem{y8}
M.~{Ricotti}, A.~{Gould},     
%  {\it A New Probe of Dark Matter and High-Energy Universe Using Microlensing}, 
{\em Astrophys. J.} {\bf 707} (2009) 979.
%[\href{http://xxx.lanl.gov/abs/astro-ph/0908.0735}{{\tt 0908.0735}}].
[arXiv:0908.0735]
\bibitem{y10}
Y.~{Yang}, X.~{Chen}, L.~{Tan}, H.~{Zong},     
%  {\it The Abundance of New Kind of Dark Matter Structures}, 
{\em Eur. Phys. J. Plus} {\bf 126} (2011) 123.
%[\href{http://xxx.lanl.gov/abs/astro-ph/1112.6228}{{\tt 1112.6228}}].
[arXiv:1112.6228]

\bibitem{y9}
Y.~{Yang}, X.~{Huang}, X.~{Chen}, H.~{Zong},     
%  {\it New Constraints on Primordial Minihalo Abundance Using Cosmic Microwave Background Observations}, 
{\em Phys. Rev.} {\bf D84} (2011) 043506.
%[\href{http://xxx.lanl.gov/abs/astro-ph/1109.0156}{{\tt 1109.0156}}].
[arXiv:1109.0156]

\bibitem{y11}
D.~{Zhang},     
 % {\it Impact of primordial ultracompact minihaloes on the intergalactic medium and first structure formation}, 
{\em MNRAS} {\bf 418} (2011) 1850.
%[\href{http://xxx.lanl.gov/abs/astro-ph/1011.1935}{{\tt 1011.1935}}].
[arXiv:1011.1935]

\bibitem{y14}
A.~S.~{Josan}, A.~M.~{Green},  
 % {\it Gamma rays from ultracompact minihalos: Potential constraints on the primordial curvature perturbation}, 
{\em Phys. Rev.} {\bf D82} (2010) 083527.
%[\href{http://xxx.lanl.gov/abs/astro-ph/1006.4970}{{\tt 1006.4970}}].
[arXiv:1006.4970]

\bibitem{y12}
P.~{Scott}, S.~{Sivertsson},    
 % {\it Gamma Rays from Ultracompact Primordial Dark Matter Minihalos}, 
{\em Phys. Rev. Lett.} {\bf 103} (2009) 211301.
%[\href{http://xxx.lanl.gov/abs/astro-ph/0908.4082}{{\tt 0908.4082}}].
[arXiv:0908.4082]

\bibitem{y13}
Y.~{Yang}, L.~{Feng}, X.~{Huang}, X.~{Chen}, L.~{Tan}, H.~{Zong},     
 % {\it Constraints on ultracompact minihalos from extragalactic γ-ray background}, 
{\em JCAP} {\bf 12} (2011) 020.
%[\href{http://xxx.lanl.gov/abs/astro-ph/1112.6229}{{\tt 1112.6229}}].
[arXiv:1112.6229]

\bibitem{radio_1}
M.~{Seffert} et al., 
[arXiv:0901.0559]

\bibitem{radio_2}
M.~{Gervasi} et al., 
{\em Astrophys. J.} {\bf 682} (2008) 223.

\bibitem{y15}
J.~{Singal} et al,      
 % {\it The ARCADE 2 Instrument}, 
{\em Astrophys. J.} {\bf 730} (2011) 138.
%[\href{http://xxx.lanl.gov/abs/astro-ph/0901.0546}{{\tt 0901.0546}}].
[arXiv:0901.0546]

\bibitem{radio3}
R.~S.~{Roger}, C.~H.~{Costain}, T.~L.~{Landecker}, C.~M.~{Swerdlyk},      
 % {\it The radio emission from the Galaxy at 22 MHz}, 
{\em Astrophys. J. Suppl.} {\bf 137} (1999) 7.
%[\href{http://xxx.lanl.gov/abs/astro-ph/9902213}{{\tt astro-ph/9902213}}].
[astro-ph/9902213]

\bibitem{radio4}
A.~E.~{Guzman}, J.~{May}, H.~{Alvarez}, K.~{Maeda},
 % {\it All-sky Galactic radiation at 45 MHz and spectral index between 45 and 408 MHz}, 
{\em A\&A} {\bf 525} (2011) A138.
%[\href{http://xxx.lanl.gov/abs/astro-ph/1011.4298}{{\tt 1011.4298}}].
[arXiv:1011.4298]

\bibitem{radio1}
C.~G.~T~{Haslam} et al, 
 % {\it A 408 MHz all-sky continuum survey. I - Observations at southern declinations and for the North Polar region}, 
{\em A\&A} {\bf 100} (1981) 209.

\bibitem{radio2}
P.~{Reich}, W.~{Reich}, 
 % {\it A radio continuum survey of the northern sky at 1420 MHz. II}, 
{\em A\&AS} {\bf 63} (1986) 205.

\bibitem{y16}
J.~{Singal} et al,     
 % {\it Sources of the radio background considered}, 
{\em MNRAS} {\bf 409} (2010) 1172.
%[\href{http://xxx.lanl.gov/abs/astro-ph/0909.1997}{{\tt 0909.1997}}].
[arXiv:0909.1997]

\bibitem{y17}
T.~{vernstrom}, D.~{Scott}, J.~V.~{Wall},      
 % {\it Contribution to the diffuse radio background from extragalactic radio sources}, 
{\em MNRAS} {\bf 415} (2011) 3641.
%[\href{http://xxx.lanl.gov/abs/astro-ph/1102.0814}{{\tt 1102.0814}}].
[arXiv:1102.0814]

\bibitem{y18}
N.~{Fornengo}, R.~{Lineros}, M.~{Regis}, M.~{Taoso},
 % {\it Possibility of a Dark Matter Interpretation for the Excess in Isotropic Radio Emission Reported by ARCADE}, 
{\em Phys. Rev. Lett.} {\bf 107} (2011) 271302.
%[\href{http://xxx.lanl.gov/abs/astro-ph/1108.0569}{{\tt 1108.0569}}].
[arXiv:1108.0569]

\bibitem{y19}
N.~{Fornengo}, R.~{Lineros}, M.~{Regis}, M.~{Taoso},
 % {\it Cosmological Radio Emission induced by WIMP Dark Matter}, 
{\em JCAP} {\bf 03} (2012) 033.
%[\href{http://xxx.lanl.gov/abs/astro-ph/1112.4517}{{\tt 1112.4517}}].
[arXiv:1112.4517]

\bibitem{y20}
D.~{Hooper}, A.~V.~{Belikov}, T.~E.~{Jeltema}, T.~{Linden}, S.~{Profumo}, T.~R.~{Slatyer}, 
{\em Phys. Rev.} {\bf D86} (2012) 103003.  
 % {\it The Isotropic Radio Background and Annihilating Dark Matter},
%[\href{http://xxx.lanl.gov/abs/astro-ph/1203.3547}{{\tt 1203.3547}}].
[arXiv:1203.3547]

\bibitem{y21}
T.~{Bringmann}, P.~{Scott}, Y.~{Akrami},  
 % {\it Improved constraints on the primordial power spectrum at small scales from ultracompact minihalos}, 
{\em Phys. Rev.} {\bf D85} (2012) 125027.
%[\href{http://xxx.lanl.gov/abs/astro-ph/1110.2484}{{\tt 1110.2484}}].
[arXiv:1110.2484]

%\bibitem{ucmhs_clus}
%J.~R.~{Chisholm}, 
 % {\it Clustering of primordial black holes: Basic results}, 
%{\em Phys. Rev.} {\bf D73} (2006) 083504.
%[\href{http://xxx.lanl.gov/abs/astro-ph/0509141}{{\tt astro-th/0509141}}].
%[astro-ph/0509141]

\bibitem{y23}
G.~{Bertone}, D.~{Hooper}, J.~{Silk}  
 % {\it Particle Dark Matter: Evidence, Candidates and Constraints}, 
{\em Phys. Rep.} {\bf 405} (2005) 279-390.
%[\href{http://xxx.lanl.gov/abs/hep-ph/0404175}{{\tt hep-ph/0404175}}].
[hep-ph/0404175]

\bibitem{y22}
G.~{Jungman}, M.~{Kamionkowski}, K.~{Griest},   
 % {\it Gamma rays from ultracompact minihalos: Potential constraints on the primordial curvature perturbation}, 
{\em Phys. Rep.} {\bf 267} (1996) 195.
%[\href{http://xxx.lanl.gov/abs/hep-ph/9506380}{{\tt hep-ph/9506380}}].
[hep-ph/9506380]]

\bibitem{y24}
S.~{Colafranceso}, S.~{Profumo}, P.~{Ullio},
  %{\it Multi-frequency analysis of neutralino dark matter annihilations in the Coma cluster}, 
{\em A\&A} {\bf 455} (2006) 21.
%[\href{http://xxx.lanl.gov/abs/astro-ph/0507575}{{\tt astro-th/0507575}}].
[astro-ph/0507575]

\bibitem{y25}
S.~{Colafrancesco}, S.~{Profumo}, P.~{Ullio},
  %{\it Detecting dark matter WIMPs in the Draco dwarf: a multi-wavelength perspective}, 
{\em Phys. Rev.} {\bf D75} (2007) 023513.
%[\href{http://xxx.lanl.gov/abs/astro-ph/0607073}{{\tt astro-th/0607073}}].
[astro-ph/0607073]

%\bibitem{y26}
%G.~{Servant}, T.~M.~{Tait},    
 % {\it Is the Lightest Kaluza-Klein Particle a Viable Dark Matter Candidate?}, 
%{\em Nucl. Phys.} {\bf B650} (2003) 391,
%[\href{http://xxx.lanl.gov/abs/hep-ph/0206071}{{\tt hep-ph/0206071}}].

%\bibitem{y27}
%H.~C.~{Cheng}, K.~T.~{Matchev}, M.~{Schmaltz}, 
% % {\it Radiative corrections to Kaluza-Klein masses}, 
%{\em Phys. Rev.} {\bf D66} (2002) 036005,
%[\href{http://xxx.lanl.gov/abs/hep-ph/0204342}{{\tt hep-th/0204342}}].

%\bibitem{kk_cro}
%A.~E. {Baltz}, D.~{Hooper} 
% % {\it New Measurement of the Antiproton-to-Proton Flux Ratio up to 100 GeV in the Cosmic Radiation}, 
%[\href{http://xxx.lanl.gov/abs/hep-ph/0411053}{{\tt hep-ph/0411053}}].

\bibitem{ics}
S.~{Profumo}, T.~E~{Jeltema}, 
 % {\it Extragalactic Inverse Compton Light from Dark Matter Annihilation and the Pamela Positron Excess}, 
{\em JCAP} {\bf 0907} (2009) 020.
%[\href{http://xxx.lanl.gov/abs/astro-ph/0906.0001}{{\tt 0906.0001}}].
[arXiv:0906.0001]

\bibitem{magnetic_1}
P.~P~{Kronberg}, 
{\em AIPC} {\bf 558} (2001) 451.

\bibitem{magnetic_2}
T.~{Akahori}, D.~{Ryu}, 
{\em Astrophys. J.} {\bf 723} (2010) 476.
[arXiv:1009.0570]

\bibitem{pamela}
O.~{Adriani} et al, 
 % {\it New Measurement of the Antiproton-to-Proton Flux Ratio up to 100 GeV in the Cosmic Radiation}, 
%[\href{http://xxx.lanl.gov/abs/astro-ph/0810.4994}{{\tt 0810.4994}}].
{\em Phys. Rev. Lett.} {\bf 102} (2009) 051101.
[arXiv:0810.4994]

\bibitem{atic}
J.~{Chang} et al,
 % {\it An excess of cosmic ray electrons at energies of 300 $\sim$ 800GeV}, 
{\em Nature} {\bf 456} (2008) 362.

\bibitem{comptel}
G.~{Weidenspointner} et al,    
 % {\it The COMPTEL instrumental line background}, 
{\em A\&A} {\bf 368} (2001) 347.
%[\href{http://xxx.lanl.gov/abs/astro-ph/0012332}{{\tt astro-ph/0012332}}].
[astro-ph/0012332]

\bibitem{fermi}
A.~A.~{Abdo} et al, 
  %{\it Spectrum of the Isotropic Diffuse Gamma-Ray Emission Derived from 
%First-Year Fermi Large Area Telescope Data}, 
%[\href{http://xxx.lanl.gov/abs/astro-ph/1002.3603}{{\tt 1002.3603}}].
{\em Phys. Rev. Lett.} {\bf 104} (2010) 101101.
[arXiv:1002.3603]

\bibitem{chandra}
R.~C.~{Hickox}, M.~{Markevitch}, 
%  {\it Resolving the Unresolved Cosmic X-ray Background in the Chandra Deep Fields}, 
{\em Astrophys. J.} {\bf 661} (2007) L117.
%[\href{http://xxx.lanl.gov/abs/astro-ph/0702556}{{\tt astro-ph/0702556}}].
[astro-ph/0702556]

\bibitem{xuelei}
X.~{Chen}, M.~{Kamionkowski}, 
%  {\it Resolving the Unresolved Cosmic X-ray Background in the Chandra Deep Fields}, 
{\em Phys. Rev.} {\bf D70} (2004) 043502.
%[\href{http://xxx.lanl.gov/abs/astro-ph/043502}{{\tt astro-ph/043502}}].
[astro-ph/043502]

\end{thebibliography}\endgroup

%\bibitem[\protect\citeauthoryear{DarkSUSY}{2009}]{darksusy}
%http://www.physto.se/~edsjo/darksusy/
%\end{thebibliography}

%\bsp

%\label{lastpage}

\end{document}